\documentclass[letterpaper, 10 pt, conference]{ieeeconf} 
\IEEEoverridecommandlockouts
\usepackage{graphics}
\usepackage{amsmath}
\usepackage{amssymb}
\usepackage{tikz}
\usetikzlibrary{positioning,arrows.meta,calc,backgrounds}
\usepackage{xcolor}
\usepackage{float}
\usepackage{booktabs}
\usepackage{siunitx}

\title{\LARGE \bf
Cross-Scenario Controller Parameter Transformation from Real-World Driving Data through Flow Matching}

\author{Christian Zink$^{1,2}$, Maik Pfefferkorn$^{1}$, Eric Lenz$^{1}$, Felix Kallmeyer$^{2}$ and Rolf Findeisen$^{1}$
\thanks{$^{1}$C. Zink, M. Pfefferkorn, E. Lenz and R. Findeisen are with the Control and Cyber-Physical Systems Laboratory,
        Technical University of Darmstadt, Darmstadt, Germany, email: {\{rolf.findeisen, maik.pfefferkorn, elenz\}@iat.tu-darmstadt.de}}
\thanks{$^{2}$C. Zink and F. Kallmeyer are with Driver Assistance Systems Predevelopment, Volkswagen AG, Wolfsburg, Germany, email:
        {\{christian.zink, felix.kallmeyer\}@volkswagen.de}}
\thanks{This work was supported by Volkswagen AG.}
}

\begin{document}
\definecolor{cSeq}{RGB}{230,240,255}   
\definecolor{cEnc}{RGB}{224,255,224}   
\definecolor{cAgg}{RGB}{255,224,178}   
\definecolor{cCon}{RGB}{230,240,255}   
\definecolor{cVAE}{RGB}{212,237,218}   
\definecolor{cLat}{RGB}{217, 236, 236} 
\definecolor{cDec}{RGB}{255,235,230}   
\definecolor{cLoss}{RGB}{255,249,196}  
\definecolor{cEdge}{RGB}{70,70,70}	   

\definecolor{taborange}{HTML}{f5ba3b}
\definecolor{tabblue}{HTML}{1f77b4}
\definecolor{tabpurple}{HTML}{9467bd}
\definecolor{tabgreen}{HTML}{2ca02c}
\definecolor{tred}{HTML}{ff0000}

\DeclareRobustCommand{\slowmark}{
  \tikz[baseline=-0.6ex]
  \draw[draw=taborange, line width=0.4pt, fill=taborange!40] (0,0) circle (0.45ex);
}

\DeclareRobustCommand{\fastmark}{
  \tikz[baseline=-0.6ex]
  \draw[draw=tabblue, line width=0.4pt, fill=tabblue!40] (0,0) circle (0.45ex);
}

\DeclareRobustCommand{\linearmark}{
  \tikz[baseline=-0.6ex]
  \draw[draw=tabpurple, line width=0.4pt, fill=tabpurple!60] (0,0) circle (0.45ex);
}

\DeclareRobustCommand{\linearmarkred}{
  \tikz[baseline=-0.6ex]
  \draw[draw=tred, line width=0.3pt, fill=tabpurple!60] (0,0) circle (0.45ex);
}

\DeclareRobustCommand{\fmmark}{
  \tikz[baseline=-0.6ex]
  \draw[draw=tabgreen, line width=0.4pt, fill=tabgreen!60] (0,0) circle (0.45ex);
}

\DeclareRobustCommand{\fmmarkred}{
  \tikz[baseline=-0.6ex]
  \draw[draw=tred, line width=0.3pt, fill=tabgreen!60] (0,0) circle (0.45ex);
}

\newcommand{\slowline}{\textcolor{taborange!70}{\rule[0.5ex]{1.5em}{0.5pt}}}
\newcommand{\fastline}{\textcolor{tabblue!70}{\rule[0.5ex]{1.5em}{0.5pt}}}
\newcommand{\linearline}{\textcolor{tabpurple!70}{\rule[0.5ex]{1.5em}{0.5pt}}}
\newcommand{\fmline}{\textcolor{tabgreen!70}{\rule[0.5ex]{1.5em}{0.5pt}}}

\newcommand{\tbox}[1]{
	\tikz[baseline=(X.base)]{
		\node[draw, inner sep=1.8pt, line width=0.4pt, rectangle] (X) {#1};
	}
}

\newlength{\modelwidth}
\settowidth{\modelwidth}{FM CFG cond DrvEmb}

\maketitle

\begin{abstract}
Advanced Driver Assistance Systems have become increasingly widespread, aiming to support drivers, reduce workload, and improve traffic safety.
However, many systems are rarely used because their behavior feels unnatural and does not match individual driving behavior. 
The growing availability of large-scale, real-world human driving data enables more human-like and driver-individual assistance functions.
This paper formulates the adaptation of Adaptive Cruise Control parameters as a driver-individual scenario-to-scenario transport problem, transforming controller parameters from one driving context to another. 
Using flow matching, we train a generative model to predict the target-scenario behavior of previously unseen drivers from only a few demonstrations in a source scenario. 
We compare the approach with a linear parameter-shift baseline and evaluate flow-matching variants incorporating driver-consistent coupling, driver embeddings, and classifier-free guidance based on driving profiles extracted from trajectory data using a variational autoencoder.
\end{abstract}

\section{Introduction and Motivation}
\label{sec:introduction}
Over the past two decades, an increasing number of Advanced Driver Assistance Systems (ADAS) have been integrated into modern vehicles to support drivers, reduce workload, and improve traffic safety~\cite{musau25:afidwtaadas}.
Despite its widespread availability, Adaptive Cruise Control (ACC), which maintains velocity and distance to a preceding vehicle, is often rarely used or disabled because its behavior is perceived as unnatural and inconsistent with individual driving styles~\cite{liu21:accse,plebe24:hiad}.
In particular, drivers may expect the vehicle to behave differently from the behavior produced by the assistance function or from how they would drive themselves~\cite{chen24:tcfmfcacc,zhu19:tdsopacc}.
The growing availability of real-world human driving data provides an opportunity to develop more human-like and driver-individual assistance functions~\cite{zhao23:rtlodgfpacc,wang22:gppacc}.

Data-driven personalization has been explored across driver assistance systems, including ACC.
For example, driver-specific cost functions have been inferred using inverse reinforcement learning to re-plan longitudinal motion~\cite{ozkan21:pacc,zhao23:rtlodgfpacc}.
Other approaches classify drivers into a small number of driving-style categories and switches between pre-tuned controller sets, e.g., using Gaussian mixture models with fuzzy PID control~\cite{su18:paccdc,nava19:pacc}.
Both approaches require the target driver to be observed in each situation for which the controller is to be adapted.
Given the broad range of scenarios and environmental conditions, such coverage is impractical.
A mechanism is therefore needed to transfer observed driving characteristics between scenarios.

This work addresses this problem by learning a driver-individual \emph{scenario-to-scenario-transformation} from recorded driving data using flow matching~\cite{lipman22:fmgm}.
Specifically, we learn from a large number of drivers a vector field that transforms a distribution of ACC controller-parameter points in one scenario into the corresponding distribution in another.
Scenarios are defined by external driving conditions rather than driver-specific behavior, allowing different drivers to exhibit different behaviors under the same scenario.
These conditions may include environmental factors such as lighting, temperature, and road-surface condition, as well as road type, lane, traffic density, or the relative velocity during an approach to a preceding vehicle.

\begin{figure}[bp]
\vspace*{-0.1cm}
\begin{center}
\begin{tikzpicture}
    \node (red) at (0, 0) {\includegraphics[width=2.5cm]{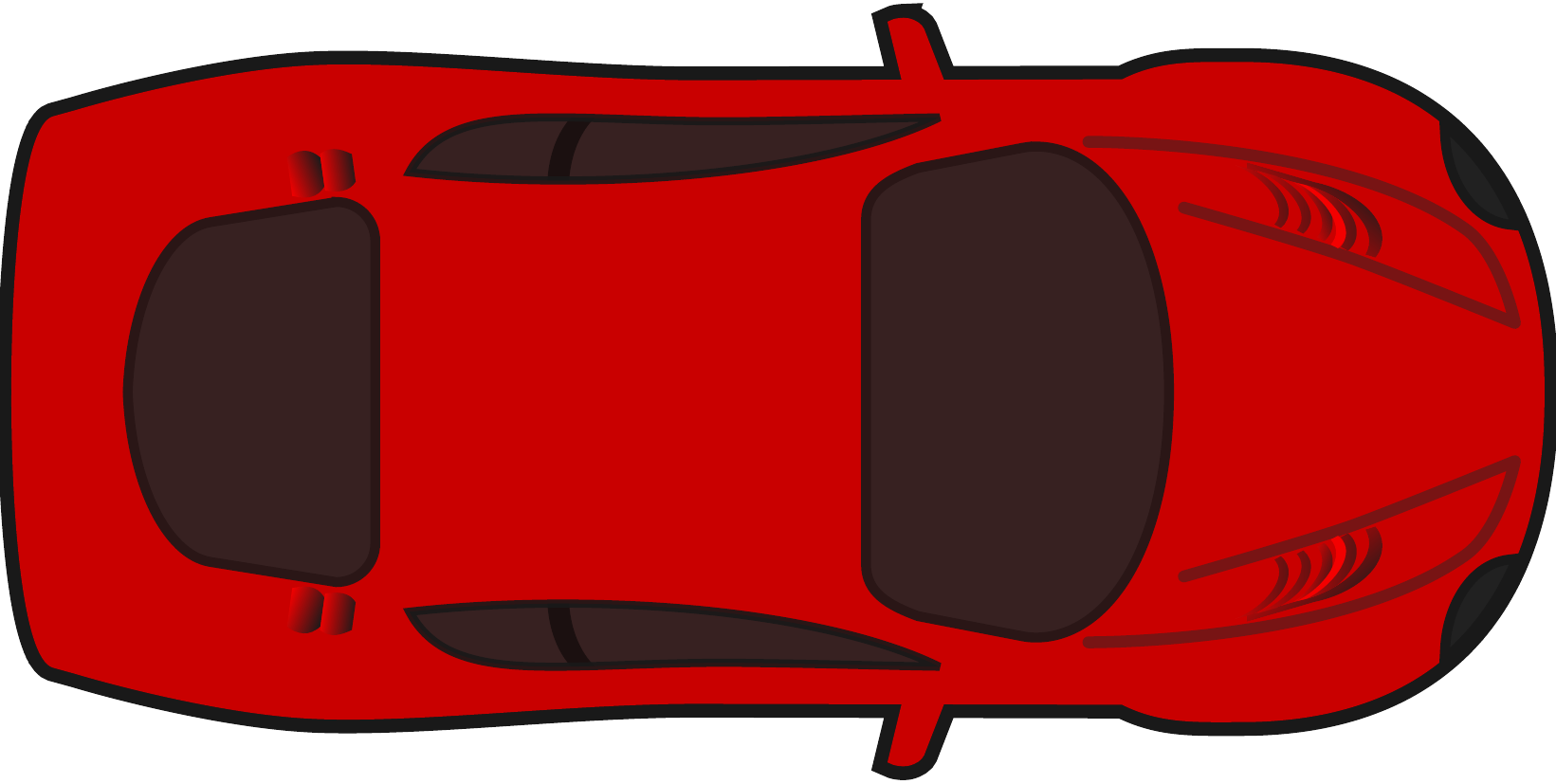}};
    \node (blue) at (5, 0) {\includegraphics[width=2.5cm]{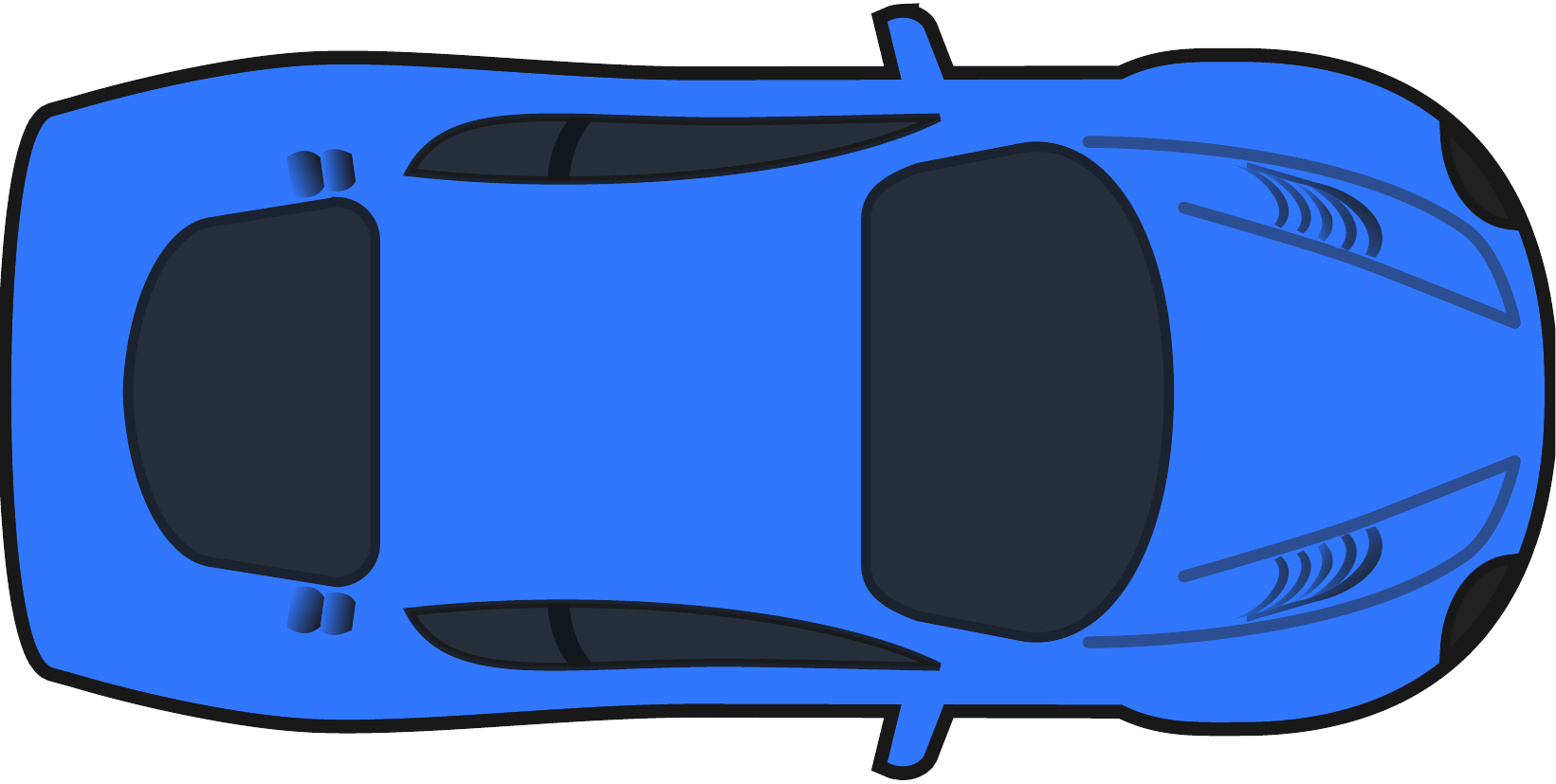}};

    \draw[->] ([xshift=-1.26mm]red.east) -- node[above=0mm] {$v$} ([xshift=0.9cm]red.east);
    \draw[->] ([xshift=-1.26mm]blue.east) -- node[above=0mm] {$v_{-1}$} ([xshift=0.9cm]blue.east);
    \draw[<->] ([xshift=-1.26mm, yshift=-8mm]red.east) -- node[below=0mm] {$D$} ([xshift=1.28mm, yshift=-8mm]blue.west);
	\draw ([xshift=-1.3mm]red.east) -- ([xshift=-1.3mm, yshift=-10mm]red.east);
	\draw ([xshift=1.32mm]blue.west) -- ([xshift=1.32mm, yshift=-10mm]blue.west);
\end{tikzpicture}
\vspace*{-0.8cm}
\caption{The ego vehicle (red) is either approaching (decreasing distance) or following (maintaining a nearly constant distance) the preceding vehicle (blue). $D$ denotes the inter-vehicle distance, while $v_{-1}$ and $v$ represent the velocities of the preceding and ego vehicle, respectively.}
\label{fig:follow-drive}
\vspace*{-0.2cm}
\end{center}
\end{figure}
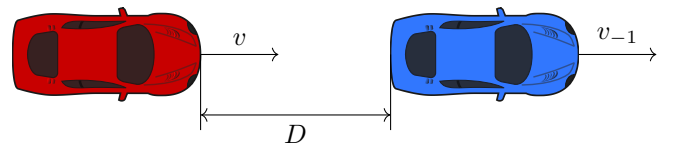

Of particular interest is the scenario-to-scenario transformation of ACC parameters between slow and fast approach maneuvers, as illustrated in Fig.~\ref{fig:follow-drive}, although the underlying problem is not limited to ACC or automated driving.
The practical benefit of this formulation is its predictive nature:
given only a few observations of a previously unseen driver in a source scenario, the learned transformation predicts how that driver would behave in a different target scenario without requiring observations in the target scenario.

Initial results on scenario-to-scenario transformation were presented in~\cite{zink26:lbccp}, where slow and fast approaches are represented by distributions of identified ACC parameter vectors and their aggregated difference defines the transformation.
This formulation reduces the transformation to a global shift in parameter space, capturing only the target distribution mean while neglecting differences in shape, variance, and orientation.
It therefore cannot describe how individual regions of the source distribution map to corresponding regions of the target distribution.

Flow matching instead transports a source distribution into a target distribution through an ordinary differential equation~\cite{lipman22:fmgm}.
While commonly used to generate data by transporting a simple Gaussian distribution to a complex target distribution, formulations with non-Gaussian source distributions~\cite{tong23:cfm,liu22:fsaf} enable direct transport between empirical distributions, as required here for measured source and target parameter sets.
Recent work in control and robotics applies flow matching to generate actions or trajectories directly~\cite{kurtz25:gpc,funk24:af}.
Here, we apply it to the parameter space of an interpretable low-order controller, providing a low-dimensional, physically meaningful transport space while retaining the closed-loop structure for analysis, simulation, and validation.

The main contributions of this work are as follows:
First, we formulate driver-individual controller parameterization as a distribution transport problem between driving scenarios and show that flow matching enables driver-individual scenario-to-scenario transformation of ACC approach parameters.
Second, we propose a transformer-based $\beta$-variational autoencoder ($\beta$-VAE) with a set-aggregator to extract compact driving profiles from raw approach trajectories~\cite{lee19:st}.
A prototype-consistency scheme aligns profiles inferred from sets of maneuvers with those inferred from individual maneuvers in a common latent space, enabling personalization from a single observed approach.
In our experiments, this alignment provides a more suitable conditioning signal than a standard autoencoder or $\beta$-VAE.
Third, on a large-scale real-world fleet dataset, we show that the proposed transport reduces the Wasserstein-2 distance to the measured target parameter distribution by roughly a factor of six compared with the linear transformation of~\cite{zink26:lbccp}.
We further compare driver-consistent coupling, driver embedding, and classifier-free guidance (CFG) based on the $\beta$-VAE driving profiles as alternative flow-matching variants~\cite{ho22:cfdg}.

The remainder of the paper is organized as follows.
Section~\ref{sec:parameter-representation} introduces the data-driven parameter representation and formulates the problem. 
Section~\ref{sec:dataset-preprocessing} describes the dataset and preprocessing used to obtain comparable ACC approach maneuvers from real-world driving data. 
Section~\ref{sec:flow-matching-vae} presents the proposed flow-matching framework for driver-individual scenario-to-scenario parameter transformation, including the investigated coupling and conditioning strategies. 
Section~\ref{sec:results} provides qualitative and quantitative evaluations and compares the proposed methods with the linear transformation baseline.
Finally, Section~\ref{sec:conclusion} summarizes the main findings and discusses future research directions.

\section{Problem Formulation}
\label{sec:parameter-representation}

This work considers the transformation of ACC controller parameters between driving scenarios to reproduce driver-individual behavior during approach maneuvers. 
We use a proportional–derivative (PD) control structure, which is widely used in ACC due to its low computational cost and robustness~\cite{lefevre15:dmpds}. 
Since the relevant dynamics of an approach maneuver are longitudinal, only the longitudinal vehicle dynamics are considered. 
These are modeled as a kinematic integrator chain,
\begin{equation}
    \begin{bmatrix}\dot v(t) \\ \dot D(t) \end{bmatrix} =
    \begin{bmatrix} 0 & 0 \\ -1 & 0 \end{bmatrix} \begin{bmatrix} v(t) \\ D(t) \end{bmatrix} +
    \begin{bmatrix} 1 \\ 0 \end{bmatrix} a(t) +
    \begin{bmatrix} 0 \\ 1 \end{bmatrix} v_{-1}(t),
    \label{eq:plant}
\end{equation}
where $v(t)$ and $a(t)$ denote the ego vehicle's velocity and acceleration, respectively, $D(t)$ is the relative distance to the preceding vehicle, and $v_{-1}(t)$ is its velocity.

The ACC acceleration $a_\mathrm{ACC}(t)$ is
\begin{equation}
    a_\mathrm{ACC}(t) = K_\mathrm{P} \cdot e_\mathrm{D}(t) + K_\mathrm{D} \cdot e_{v_{r}}(t),
\end{equation}
where $e_\mathrm{D}(t)= D(t) - D_\mathrm{\min}(t)$, $e_{v_\mathrm{r}}(t)= v_{-1}(t) - v(t)$, and $D_\mathrm{\min}(t) = v(t) \cdot \mathrm{THW_{set}}$.
Thus, each approach maneuver is represented by the controller parameter vector $\theta=[K_\mathrm{P},K_\mathrm{D}]^\top$.

The objective is to learn, from human driving data, how $\theta$ changes between driving scenarios while preserving driver-specific behavior. 
In particular, we consider slow and fast approach maneuvers, distinguished by their initial relative velocity $v_{\mathrm{r},0}$. 
The resulting parameter distributions form the source and target distributions for the scenario-to-scenario transformation studied in this work.

\section{Dataset and Data Preprocessing}
\label{sec:dataset-preprocessing}
The data used in this study originates from a large-scale field dataset collected by the Volkswagen Group during regular, real-world vehicle operation rather than under a controlled experimental protocol.
The dataset therefore captures a naturalistic distribution of human driving behavior across divers environments and traffic conditions.

In total, the dataset comprises over $16{,}000$ measurements from more than $170$ internal drivers operating a fleet of around $30$ vehicles spanning three distinct vehicle models, all equipped with combustion engines.
The measurements were collected over a two-year period in over $30$ countries, providing substantial diversity in drivers, road types, and geographic regions.
Each measurement is associated with an anonymous driver identifier, enabling driver-specific analysis and maneuver aggregation.

The vehicles were not equipped with an ACC or other driver assistance systems.
Thus, all approach maneuvers analyzed in this work represent purely manual driving.
The perception of the preceding vehicle relies on a mid-range radar sensor in combination with a front camera.
The radar detects all relevant objects in front of the ego vehicle and measures their relative velocity $v_\mathrm{r}$, while the front-camera identifies and confirms the preceding vehicle.

\subsubsection{Preprocessing}
The raw recordings were filtered and normalized to obtain comparable approach maneuvers.
For reproducibility, only maneuvers in which the ego vehicle traveled at $v \geq \SI{20}{\m \per \s}$ when the preceding vehicle was detected were retained.
The preceding vehicle was required to travel at approximately constant velocity, deviating by at most $\pm \SI{2}{\m \per \s}$ from its mean.
All maneuvers were initialized at a common relative distance of $D = 120\,\mathrm{m}$, beyond which no measurement exhibited a significant change in ego velocity, thereby effectively fixing the initial distance $D_0$.

All signals were resampled by linear interpolation at $\SI{10}{Hz}$.
To suppress high-frequency vibrations and sensor artifacts while preserving the maneuver dynamics, the ego vehicle's velocity $v(t)$, acceleration $a(t)$, and the relative velocity $v_\mathrm{r}(t)=v_{-1}(t)-v(t)=\frac{d}{dt}D(t)$ were smoothed using a fourth-order Butterworth low-pass filter with a $\SI{2}{\Hz}$ cutoff, applied in both forward and reverse directions to avoid phase distortion.
Here, $v_{-1}(t)$ denotes the preceding-vehicle velocity and $D(t)$ the inter-vehicle distance.
The distance $D(t)$ was obtained by integrating the filtered relative velocity $v_\mathrm{r}(t)$, from which $v_{-1}(t)$ and the time headway $\mathrm{THW}(t)=D(t)/v(t)$ were subsequently computed.

\subsubsection{Filtering}
Following the general preprocessing, three additional filtering criteria were applied to isolate representative, driver-consistent approach maneuvers.

First, the desired time headway was constrained.
A steady-state time headway below $0.9\,\mathrm{s}$ already constitutes a violation of German traffic regulations~\cite{filzek02:avaa}, while gaps exceeding $3\,\mathrm{s}$ are generally not considered representative of car-following behavior~\cite{hulst99:aatac,risto13:tas}.
The desired time headway $\mathrm{THW_{set}}$ was defined as the actual time headway at the end of each measurement, once steady following was reached, yielding $\mathcal{D}_\mathrm{THW} = \mathcal{D}(0.9\,\mathrm{s}\leq \mathrm{THW_{set}} \leq3\,\mathrm{s})$,
where $\mathcal{D}(\cdot)$ denotes the subset of $\mathcal{D}$ satisfying the stated condition.

Second, the approach behavior --- in particular its temporal characteristics and deceleration gradient --- is strongly governed by the initial relative velocity $v_\mathrm{r,0}$.
To ensure initial comparable conditions, it was restricted to 
\begin{equation}
	\mathcal{D}_{\mathrm{THW} \cap \mathrm{v_{r}}} = \mathcal{D}_\mathrm{THW}(-\SI{12}{\m \per \s}\leq v_\mathrm{r,0}\leq-\SI{4}{\m \per \s})\;.
\end{equation}
The lower bound $v_\mathrm{r,0}=-\SI{12}{\m \per \s}$ corresponds to the minimum observed value, while the upper bound $v_\mathrm{r,0}=-\SI{4}{\m \per \s}$ represents the maximum relative velocity at which active driver intervention becomes necessary. 

Finally, to extract driver-specific profiles from approach behavior, only the six drivers contributing the largest number of maneuvers were retained. 
This resulted in $T_d = 83, 45, 41, 33, 35,$ and $34$ maneuvers for $d = 0, \dots, 5$, respectively.

\subsubsection{Scenario-Based Parameter Identification}
For the scenario-to-scenario transformation, these maneuvers were divided according to the initial relative velocity $v_\mathrm{r,0}$ into slow approaches ($\SI{-5}{\m\per\s} \leq v_\mathrm{r,0} \leq \SI{-4}{\m\per\s}$) and fast approaches ($\SI{-12}{\m\per\s} \leq v_\mathrm{r,0} \leq \SI{-8}{\m\per\s}$). 
The resulting subsets define the slow and fast approach scenarios considered throughout the remainder of the paper.

For each approach trajectory, the controller parameter vector $\theta=[K_\mathrm{P},K_\mathrm{D}]^\top$ was then identified by minimizing the deviation between the simulated closed-loop response and the measured trajectory. 
The resulting nonlinear least-squares problem was solved numerically using a quasi-Newton method. 
The identified parameter vectors form the corresponding point clouds in the PD parameter space.

\section{Scenario-to-Scenario Parameter Transformation using Flow Matching}
\label{sec:flow-matching-vae}
This section uses flow matching to transform the controller-parameter distributions between driving scenarios. The identified parameter vectors form point clouds for slow and fast approaches, serving as the source and target distributions, respectively.
Following the general formulation, three variants are introduced that differ in how the driving style enters the model: coupling restricts the pairing of source and target samples during training, while driver embedding and CFG condition the vector field on the driver identity or a $\beta$-VAE driving-profile representation, respectively.
As coupling and conditioning act on different parts of the model, both may be combined.

\subsection{Flow Matching}
\label{subsec:fm}
Flow Matching transforms a source distribution $p_\mathrm{init}$ into a complex target distribution $p_\mathrm{data}$ by means of an ordinary differential equation of the form
\begin{equation}
    \frac{d}{dt} x_t =  u_t^{\phi}(x_t), \quad t\in[0,1],
	\label{eq:fm_ode}
\end{equation}
which describes the transport of probability mass from the source to the target distribution.
Here, $u_t^{\phi}$ denotes a time-dependent vector field parameterized by a neural network with weights $\phi$, while $x_t$ denotes the state of a sample at time $t$ along the transport trajectory.
This trajectory is described by a probability path that linearly interpolates between an initial sample $x_0$ and a target sample $x_1$ according to
\begin{equation}
 	x_t = x_1\,t + (1-t)\,x_0, \quad (x_0,x_1) \sim \pi, \quad t \in [0,1].
 	\label{eq:fm_interpol_path}
\end{equation}
Flow Matching is commonly used to learn a transport from a simple noise distribution to a complex data distribution, thereby enabling the generation of samples from the latter.
In contrast to this common noise-to-data setting, here $p_\mathrm{init}$ and $p_\mathrm{data}$ are the parameter distributions of two driving scenarios: $p_\mathrm{init}$ collects the optimal controller parameters $x_0 \equiv \theta  = [K_\mathrm{P}, K_\mathrm{D}]^\top$ of the slow-approach scenario and $p_\mathrm{data}$ those of the fast-approach scenario, denoted $x_1$.

The model is trained with the conditional flow matching objective
\begin{equation}
	\mathcal{L}_\mathrm{FM}(\phi) =
	\mathbb{E}_{t,\: (x_0,\: x_1)\sim\pi} \left[\|u_t^{\phi}(x_t)-(x_1-x_0)\|_2^2\right],
	\label{eq:fm_loss}
\end{equation}
with $t \sim \mathcal{U}(0,1)$~\cite{holderrieth25:fmdm}.
At inference, a sample is generated by integrating \eqref{eq:fm_ode} from $t=0$ to $t=1$ using a numerical solver, such as the explicit Euler method.
Starting from $x_0 \sim p_\mathrm{init}$, the learned vector field $u_t^{\phi}(x_t)$ transports the sample toward the target distribution $p_\mathrm{data}$.

\subsubsection{Coupling}
\label{subsec:coup_drv}
A natural way to introduce driver-specific information without changing the network architecture is to restrict which source and target samples may be paired during training~\cite{albergo24:siwddc}.
To formalize this, let $d \in \{0,\dots,N_d-1\}$, where $N_d$ is the number of drivers, denote the driver identity and let $p_\mathrm{init}^{(d)}$, $p_\mathrm{data}^{(d)}$ be the corresponding driver-specific slow- and fast-approach parameter distributions.
Each driver $d$ contributes $n_d^\mathrm{init}$ slow- and $n_d^\mathrm{data}$ fast-approach samples, with totals $N_\mathrm{init} = \sum_d n_d^\mathrm{init}$ and $N_\mathrm{data} = \sum_d n_d^\mathrm{data}$.
In this work, we consider two variants of the coupling $\pi$ in~\eqref{eq:fm_interpol_path} and \eqref{eq:fm_loss}, while keeping the network inputs unchanged.

In \emph{pooled coupling}, the source and target are drawn independently from the population marginals, resulting in
\begin{equation}
	\pi_{\mathrm{pool}}(x_0,x_1) = p_\mathrm{init}(x_0)\,p_\mathrm{data}(x_1).
	\label{eq:coup_pool}
\end{equation}
A training pair may therefore combine the slow-approach sample of one driver with the fast-approach sample of a different driver, so that the learned vector field captures only the population-level transport.

In \emph{driver-consistent coupling}, both samples are constrained to originate from the same driver, whereby the driver is selected uniformly from $d \sim \mathcal{U}\{0,\dots,N_d-1\}$ with $p(d)=1/N_d$.
This yields
\begin{equation}
	\pi_{\mathrm{drv}}(x_0,x_1) =
	\frac{1}{N_d} \sum_{d=0}^{N_d-1} p_\mathrm{init}^{(d)}(x_0)\, p_\mathrm{data}^{(d)}(x_1).
\label{eq:coup_drv}
\end{equation}
The driver-consistent coupling (DCC)~\eqref{eq:coup_drv} deliberately weights all drivers equally, whereas the pooled coupling~\eqref{eq:coup_pool} uses data-proportional weighting.
This ensures that drivers with fewer recorded approaches still contribute equally to the driver-specific transport and are not dominated by drivers with a larger number of samples, which avoids a bias toward drivers with many recordings.
However, both variants act solely on the sample pairing without providing the driver identity as a network input.

\subsubsection{Driver Embedding}
The vector field is therefore additionally conditioned on a learnable driver embedding $\xi_d = \mathrm{Emb}(d) \in \mathbb{R}^{8}$, which maps the nominal driver index $d$ to a trainable vector.
Unlike a direct numerical encoding of $d$, the embedding does not impose an artificial ordering on driver identities and can learn a representation in which distances reflect behavioral similarity. 
Here, we set its dimension to $8$ to balance expressiveness and the number of additional parameters given the small number of drivers.

The vector field then reads $u_t^{\phi}(x_t,e_d)$ and is trained by minimizing the flow matching objective
\begin{equation}
\begin{split}
	\mathcal{L}_\mathrm{FM}^\mathrm{DrvEmb}(\phi)&=
	\frac{1}{N_d}\sum_{d=0}^{N_d-1}\mathbb{E}_{t,\: (x_0,\: x_1)\sim\pi_{\mathrm{drv}}^{(d)}}\\
	&\quad
	\left[\|u_t^{\phi}(x_t,e_d)-(x_1-x_0)\|_2^2\right],
\end{split}
\end{equation}
where
$\pi_{\mathrm{drv}}^{(d)}(x_0,x_1) = p_\mathrm{init}^{(d)}(x_0)\, p_\mathrm{data}^{(d)}(x_1)$ is the driver-specific component of~\eqref{eq:coup_drv}.

\subsubsection{Classifier-Free Guidance}
\label{subsec:cfg}
To steer the flow toward a specific driving profile without training a separate classifier, we employ CFG~\cite{ho22:cfdg}.
The conditioning signal is the driver profile $z \in \mathbb{R}^{d_z}$ extracted by a $\beta$-VAE, see \ Sec.~\ref{subsec:vae}.
During training, the conditioning information is randomly replaced by a learnable null token $\varnothing$ with probability $p_\mathrm{uncond}=0.4$, so that the same network learns both the conditional field $u_t^\phi(x_t \mid z)$ and the unconditional field $u_t^\phi(x_t \mid \varnothing)$ using the conditional flow-matching objective
\begin{equation}
	\mathcal{L}_\mathrm{CFG}(\phi) \! = \!
	\mathbb{E}_{t,(x_0, x_1)\sim\pi,z} \left[\|u_t^{\phi}(x_t \! \mid \! z) \! - \! (x_1 \! - \! x_0)\|_2^2\right].
\end{equation}
At inference, the two predictions are linearly extrapolated using a guidance scale $w \ge 0$,
\begin{equation}
\tilde{u}_t^\phi(x_t \mid z) =
	u_t^\phi(x_t \mid \varnothing)+ w \bigl(u_t^\phi(x_t \mid z) - u_t^\phi(x_t \mid \varnothing)\bigr),
\label{eq:cfg_guided}
\end{equation}
and $\tilde{v}_t^\phi$ replaces $u_t^\phi$ in~\eqref{eq:fm_ode}.
For $w=0$ generation is unconditional, $w=1$ recovers the purely conditional model, whereas $w>1$ amplifies the influence of the driving-profile latent variable $z$ and sharpens the generated distribution around the individual driver characteristics.
Since excessive guidance reduces sample diversity and may shift samples beyond the target distribution, we use a moderate guidance scale of $w=1.5$, which strengthens the driver-specific conditioning while preserving the spread of the target distribution.

\subsection{Extraction of the driver profile}
\label{subsec:vae}
This section describes how the driver-specific conditioning signal of the flow matching model is extracted from recorded approach trajectories.
First, the requirements for such a profile are established, and the limitations of a naive encoder are discussed.
We then present the proposed transformer-based $\beta$-VAE with set-aggregator, its components, and the training scheme that aligns set-based and single-approach profiles in a common latent space.

\subsubsection{Requirements}
The flow matching model in Sec.~\ref{subsec:cfg} is guided by a driver-specific conditioning signal.
Therefore, a compact, low-dimensional representation vector $z \in \mathbb{R}^{d_z}$ is required that encodes the individual driving style and serves as the conditioning variable of the CFG-based flow matching model.
Each approach is represented by a signal trajectory $s_i \in \mathbb{R}^{T\times C}$ comprising $T=600$ time steps (total horizon of $\SI{60}{\s}$) and the $C=7$ channels $[D(t),v_\mathrm{r}(t),a(t),\mathrm{THW}(t),\mathrm{iTTC}(t),j(t),v(t)]^\top$ and a context vector $c_i \in \mathbb{R}^{C}$ summarizing its initial engagement conditions.
Here, $\mathrm{iTTC}(t)=-v_\mathrm{r}(t)/D(t)$ denotes the inverse time-to-collision and $j(t)$ the jerk.
The context vector is obtained by averaging $s_i(t)$ over the initial $5$ samples.

Two properties are essential: (i) the profile must be driver-discriminative rather than maneuver-specific;
(ii) profiles inferred from multiple approaches and a single approach must lie in the same latent space, allowing a driver represented by multiple recorded maneuvers during training to be characterized from a single observed approach at inference.

A natural choice is to learn such a representation directly from the recorded approach trajectories using an autoencoder or a $\beta$-VAE.
In our experiments, however, neither architecture yielded useful profiles:
the latent representation was dominated by maneuver-specific characteristics, preventing trajectories of the same driver from being mapped consistently to a common region of the latent space.

\subsubsection{Overview of the architecture}
\begin{figure}[tb]
\centering
\begin{tikzpicture}[
  font=\small,
  every path/.style={draw=cEdge, line width=0.5pt},
  node distance=6mm and 8mm,
  block/.style={rectangle, rounded corners=2pt, draw=black!60, line width=0.4pt, minimum width=48mm, minimum height=8mm, align=center, inner sep=3pt},
  small/.style={rectangle, rounded corners=2pt, draw=black!60, line width=0.4pt, minimum width=20mm, minimum height=8mm, align=center, inner sep=2pt},
  lossblock/.style={rectangle, rounded corners=2pt, draw=black!60, fill=cLoss, minimum width=60mm, minimum height=8mm, align=center, inner sep=3pt, font=\footnotesize},
  flow/.style={draw=cEdge, line width=0.5pt, -{Stealth[length=2.2mm]}},
  lbl/.style={font=\small, fill=white, inner sep=1.5pt}
]

\node[block, fill=cSeq] (traj_in) {
  \textbf{Trajectories}
};

\node[block, fill=cDec, below=9mm of traj_in] (encoder) {
  \textbf{Approach-Encoder}\\[2pt]
  Seq $\rightarrow \mathrm{Emb} \in \mathbb{R}^{d_{\mathrm{model}}}$\\
  Transformer(Emb, PosEmb, CLS)
};

\node[block, fill=cAgg, below=9mm of encoder] (aggregator) {
  \textbf{Set-Aggregator}\\[2pt]
  Transformer with Self-Attention\\
  and Attention-Pooling
};

\node[block, fill=cVAE, below=9mm of aggregator] (vae) {
  \textbf{VAE Head}\\[2pt]
  $\mu$, $\log \sigma^2$, KL-Divergence \\
  $\rightarrow$ Driver-Profile in Latent Space
};

\node[small, fill=cCon, left=0.4cm of vae] (context) {
  \textbf{Context}\\[2pt]
  $c_{d,i}$
};


\node[block, fill=cDec, below=9mm of vae] (decoder) {
  \textbf{Approach-Decoder}\\[2pt]
  Transformer($c_i$, $z_{i}$, PosEmb)
};

\node[lossblock, fill=cLoss, below=9mm of decoder] (loss) {
  $\mathcal{L} = \mathcal{L}^{\mathrm{set}}_{\mathrm{rec}} + \mathcal{L}^{\mathrm{single}}_{\mathrm{rec}}
      + \beta\, \mathcal{L}_{\mathrm{KL}} + \lambda_c\, \mathcal{L}_{\mathrm{con}}$
};

\draw[flow] (traj_in)      -- node[lbl]{$s_{d,i}$}            			(encoder);
\draw[flow] (encoder)      -- node[lbl]{$z^{\mathrm{app}}_{d,i}$}       (aggregator);
\draw[flow] (aggregator)   -- node[lbl]{$h_{\mathrm{drv},d}$, $h_{d,i}$}	(vae);
\draw[flow] (vae) -- node[lbl]{$z_{\mathrm{drv},d}$, $z_{d,i}$}	(decoder);
\draw[flow] (decoder)      -- node[lbl]{$\mathcal{L}_\mathrm{rec}$} (loss);
\draw[flow] (context) 	   --										(vae);

\coordinate (inLeft) at ($(traj_in.west)$);
\draw[flow] (inLeft) -- ++(-1.4,0) -| (context.north);
\draw[flow] (context.south) |- (decoder.west);
\draw[flow] (vae.east) -- ++(1.3,0)
      node[lbl, midway, above]{$\mathcal{L}_{\mathrm{KL}}$}
      |- (loss.east);
\end{tikzpicture}
\caption{Transformer-based $\beta$-VAE with Set-Aggregator}
\label{fig:vae}
\end{figure}
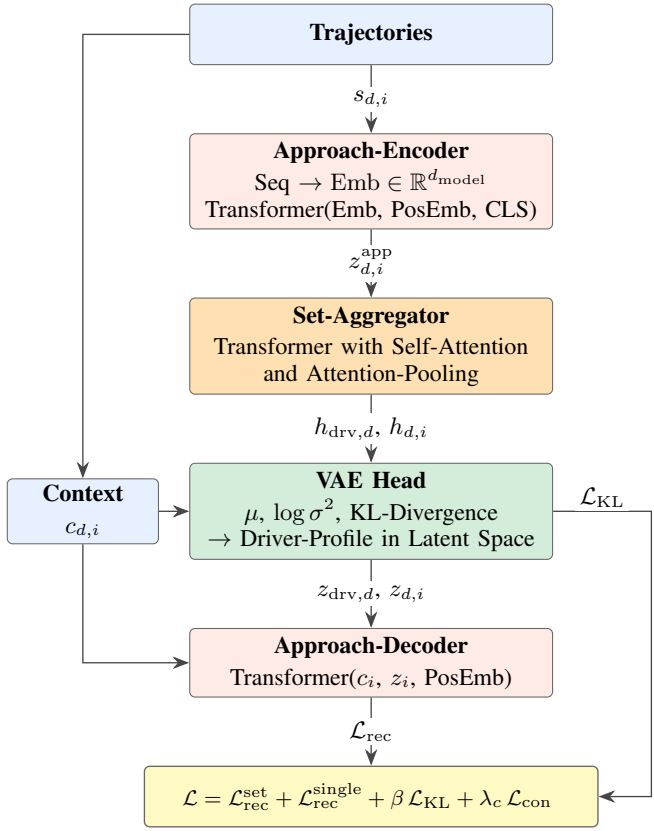

To disentangle driving style from maneuver-specific characteristics, we propose a transformer-based $\beta$-VAE with a set-aggregator, see Fig.~\ref{fig:vae}.
The model is trained as a conditional autoencoder on pairs $(s_i, c_i)$: the context $c_i$ is provided to both the VAE head and decoder but is not reconstructed, so only the trajectory $s_i$ must be recovered from the latent representation.
Thus, information already contained in $c_i$ need not be encoded in $z$, encouraging the latent representation to capture driver-specific rather than maneuver-specific characteristics.
 
The model processes each driver in four stages.
Each approach trajectory $s_i$ is first embedded individually by a shared \emph{approach-encoder}.
The resulting embeddings are fused by a \emph{set-aggregator} into a single pooled representation, which a \emph{VAE head} maps to a diagonal Gaussian posterior whose mean $\mu$ constitutes the driver profile.
Finally, an \emph{approach-decoder} reconstructs each trajectory from the latent representation, conditioned on its context vector $c_i$ containing the initial engagement conditions; the deviation between the reconstruction and the original trajectory enters the overall objective as the reconstruction loss $\mathcal{L}_\mathrm{rec}$.

The key mechanism is that the aggregator operates on two set sizes during training.
In the \emph{Set-Path}, a set of $N=20$ approaches from the same driver is aggregated into a driver-level profile capturing behavior shared across the driver's maneuvers.
In the \emph{Single-Path}, each trajectory is encoded individually ($N=1$) and its representation is encouraged to match the corresponding set-level prototype, anchoring single-approach estimates in the same driver-discriminative latent space.
This enables trajectory-specific driving profiles to serve as conditioning signal for CFG.

\subsubsection{Model components}

\paragraph{Approach-Encoder}
For each driver $d$, each approach sequence $s_{d,i}$ is embedded independently by a shared transformer encoder~\cite{vaswani17:aiayn}.
At each time step, the $C=7$ input channels are linearly projected to a $d_{\mathrm{model}}=64$-dimensional vector, augmented with learned positional embeddings, and prepended with a learnable classification token (CLS).
Three self-attention layers ($4$ heads, GELU activation) contextualize the sequence; the CLS output forms the per-approach embedding $z^{\mathrm{app}}_{d,i} \! \in \! \mathbb{R}^{d_{\mathrm{model}}}$.
Sharing the encoder across all set elements yields the set of embeddings $\{z^{\mathrm{app}}_{d,i}\}_{i=1}^{N}$.

\paragraph{Set-Aggregator}
The aggregator maps the unordered set $\{z_{d,i}^{\mathrm{app}}\}_{i=1}^{N}$ to a fixed-dimensional driver representation, suppressing maneuver-specific variation while retaining behavior shared across approaches. 
The set elements are contextualized by a two-layer transformer and pooled using a learned attention query.
The resulting representation $h_{\mathrm{drv},d}$ (or $h_{d,i}$ for a single-element set, see.\ Sec.~\ref{subsec:training}) is invariant to the ordering of the set elements.

\paragraph{VAE Head}
The pooled vector $h_d\in\{h_{\mathrm{drv},d}, h_{d,i}\}$ is concatenated with the averaged context $\bar{c}_d$ of the set and mapped to the parameters of a diagonal Gaussian posterior, $\mu_d = W_\mu[h_d,\bar{c}_d]$ and $\log\sigma^2_d = W_{\sigma}[h_d,\bar{c}_d]$, where $[\cdot,\cdot]$ denotes concatenation and $W_\mu, W_\sigma$ are learned linear maps.
The latent space driver profile is sampled using the reparameterization trick, and regularized towards a standard normal prior through the Kullback-Leibler (KL) divergence.
The posterior mean $\mu_d$ provides a deterministic driver profile used by the downstream flow matching model.

\paragraph{Approach-Decoder}
The decoder reconstructs each approach trajectory from the latent representation, conditioned on its context.
The concatenation $[c_{d,i}, z_{d,i}]$ is projected to a conditioning token that is broadcast over the horizon and added to a learned positional embedding.
Three transformer layers followed by a linear head produce the reconstructed signals $\hat{s}_{d,i} \in \mathbb{R}^{T\times C}$.

\paragraph{Training with Prototype Consistency}
\label{subsec:training}
A key requirement is that driver profiles inferred from a full set and from a single approach lie in the same latent space, allowing personalization from a single observed maneuver.
We therefore use a Set-Path--Single-Path scheme in which the approach encoder is shared and the paths differ only in their aggregation.


The two views are aligned using a prototype-consistency loss ($\mathcal{L}_{\mathrm{con}}$) that pulls each single-approach embedding towards the set-level prototype, treated as a fixed target through a stop-gradient operator $\mathrm{sg}[\cdot]$~\cite{chen21:esrl},
\begin{equation}
    \mathcal{L}_{\mathrm{con}} = \frac{1}{N}\sum_{i=1}^{N}
	\bigl\lVert \mu_{d,i} - \mathrm{sg}[\mu_{\mathrm{drv},d}]\bigr\rVert_2^2 .
\end{equation}
The stop-gradient makes the Set-Path define the target geometry, while the reconstruction and KL objectives train the aggregator on genuine multi-element sets.

\subsubsection{Latent Space and Training}
The latent space dimension was varied in a preliminary study and set to $d_z = 10$, which provided the clearest driver separation in the latent space.
Smaller values did not sufficiently separate drivers, whereas larger values increasingly encoded maneuver-specific variation without further improving separation.
The model is trained for $100$ epochs with a maximum KL-divergence weight of $\beta_{\max}=10^{-3}$ and a maximum consistency-loss weight of $\lambda_{c,\max}=0.3$, both of which are linearly annealed from zero over the first $20$ and $15$ epochs, respectively, and then kept constant.

\begin{figure}[tbp]
\begin{center}	  
\includegraphics[width=8.4cm, trim={4mm 2mm 2mm 19mm}, clip]{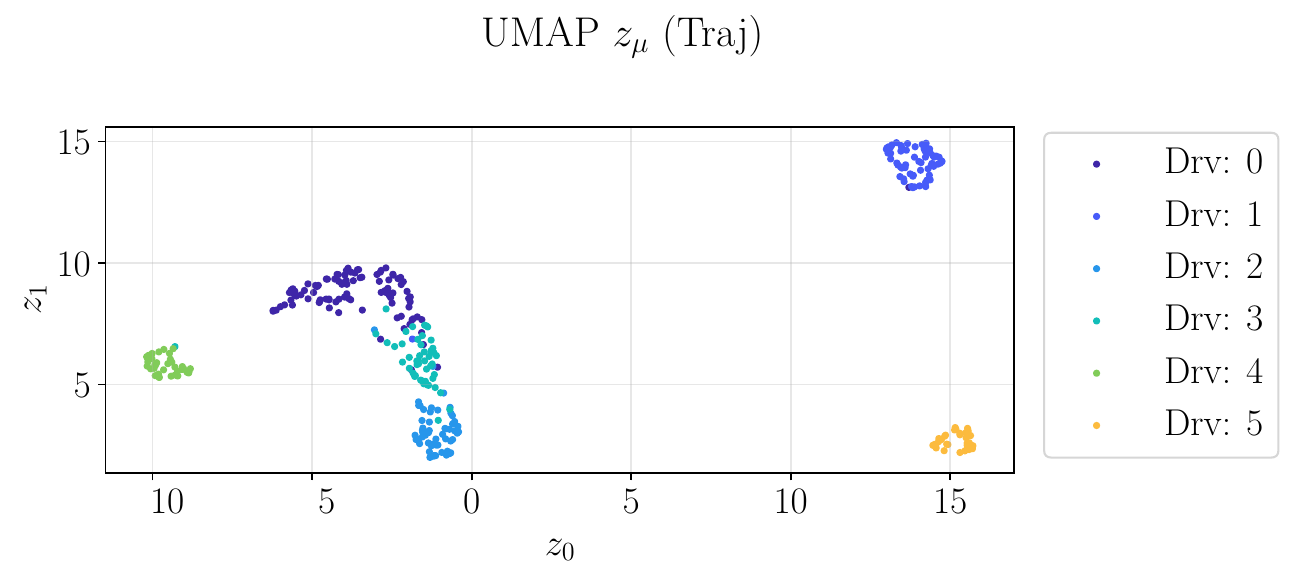}
\caption{UMAP projection of the latent space reparameterization from the per-trajectory profiles.}
\vspace{-15pt}
\label{fig:UMAP-traj}
\end{center}
\end{figure}

At inference, the driver profile is obtained deterministically from the posterior mean $\mu$, both for full sets and individual trajectories.
Encoded from single trajectories, these profiles are clearly driver-discriminative, with a 5-fold random-forest accuracy of 0.93 and a silhouette score of 0.42, see the UMAP projection in Fig.~\ref{fig:UMAP-traj}.
The consistent separation of set-based and per-trajectory profiles indicates that the Set-Path--Single-Path prototype-consistency training maps both estimates to a common, driver-discriminative latent space.

\section{Evaluation and Results}
\label{sec:results}
The evaluation follows the three contributions stated in Sec.~\ref{sec:introduction}.
First, we show qualitatively that flow matching is a suitable method for the driver-individual scenario-to-scenario transformation (Sec.~\ref{subsec:quali}).
Then, we demonstrate quantitatively that flow matching outperforms the linear transformation baseline of~\cite{zink26:lbccp} and assess the performance of the proposed flow-matching variants for the considered problem setting (Sec.~\ref{subsec:quant}).
Since all drivers are included in training, these results are in-sample and quantify how well each method represents the transformation.
Finally, we analyze the generalization capabilities of the flow-matching models to previously unseen drivers, which corresponds to the intended application (Sec.~\ref{subsec:unseen-drivers}).

\subsection{Applicability of Flow Matching}
\label{subsec:quali}
\begin{figure}[tb]
\begin{center}	  
\includegraphics[width=8.4cm, trim={0mm 0mm 0mm 0mm}, clip]{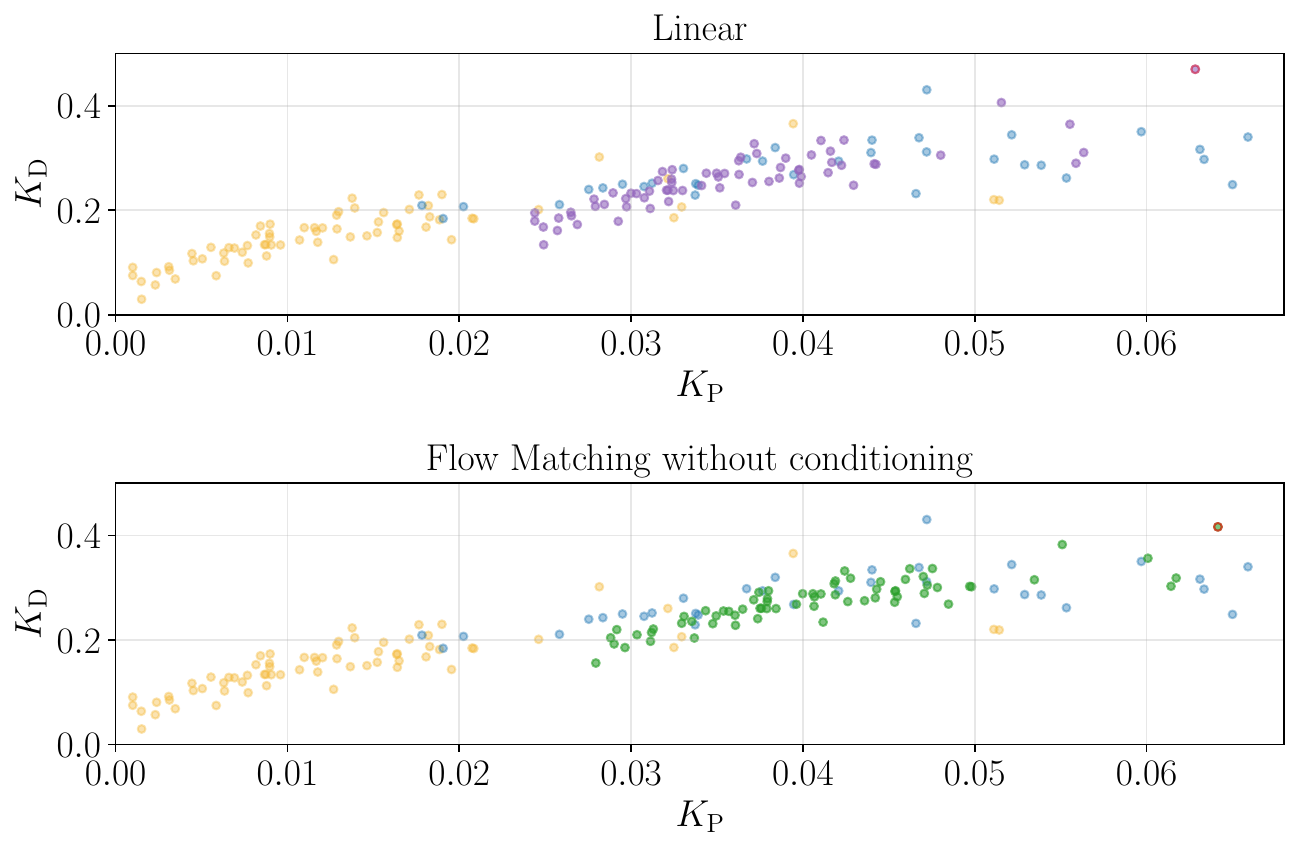}
\vspace{-5pt}
\caption{Parameter distributions ($K_\mathrm{P}, K_\mathrm{D}$) for 
\slowmark\ $\mathcal{D}_\mathrm{slow}$ and 
\fastmark\ $\mathcal{D}_\mathrm{fast}$ approach maneuvers and the generated parameter distributions 
\linearmark\ $\mathcal{D}_{\mathrm{\Delta}}$ and 
\fmmark\ $\mathcal{D}_{\mathrm{FM}}$. 
\linearmarkred\ $\mathcal{D}_{\mathrm{\Delta}}^{(m=0)}$ and 
\fmmarkred\ $\mathcal{D}_{\mathrm{FM}}^{(m=0)}$ are the analyzed parameter points.}
\label{fig:par-drv}
\end{center}
\end{figure}

Fig. \ref{fig:par-drv} compares the parameter distributions obtained using the linear transformation and flow matching.
Starting from the slow-approach dataset $\mathcal{D}_\mathrm{slow}$, both the linear transformation $\mathcal{D}_\Delta$ and flow matching $\mathcal{D}_\mathrm{FM}$ aim to reproduce the measured fast-approach dataset $\mathcal{D}_\mathrm{fast}$.
The linear transformation rigidly shifts $\mathcal{D}_\mathrm{slow}$ by $\Delta$ and therefore preserves its size and orientation, so that the transported points only partially cover the target region, whose spread differs from that of the source distribution.
Flow matching, in contrast, treats each region as a distribution and transports it through the learned vector field~\eqref{eq:fm_ode}, so that $\mathcal{D}_\mathrm{FM}$ more closely overlaps $\mathcal{D}_\mathrm{fast}$ in both location and shape.
%

\begin{figure}[tb]
\begin{center}	  
\includegraphics[width=8.4cm, trim={12mm 12mm 12mm 12mm}, clip]{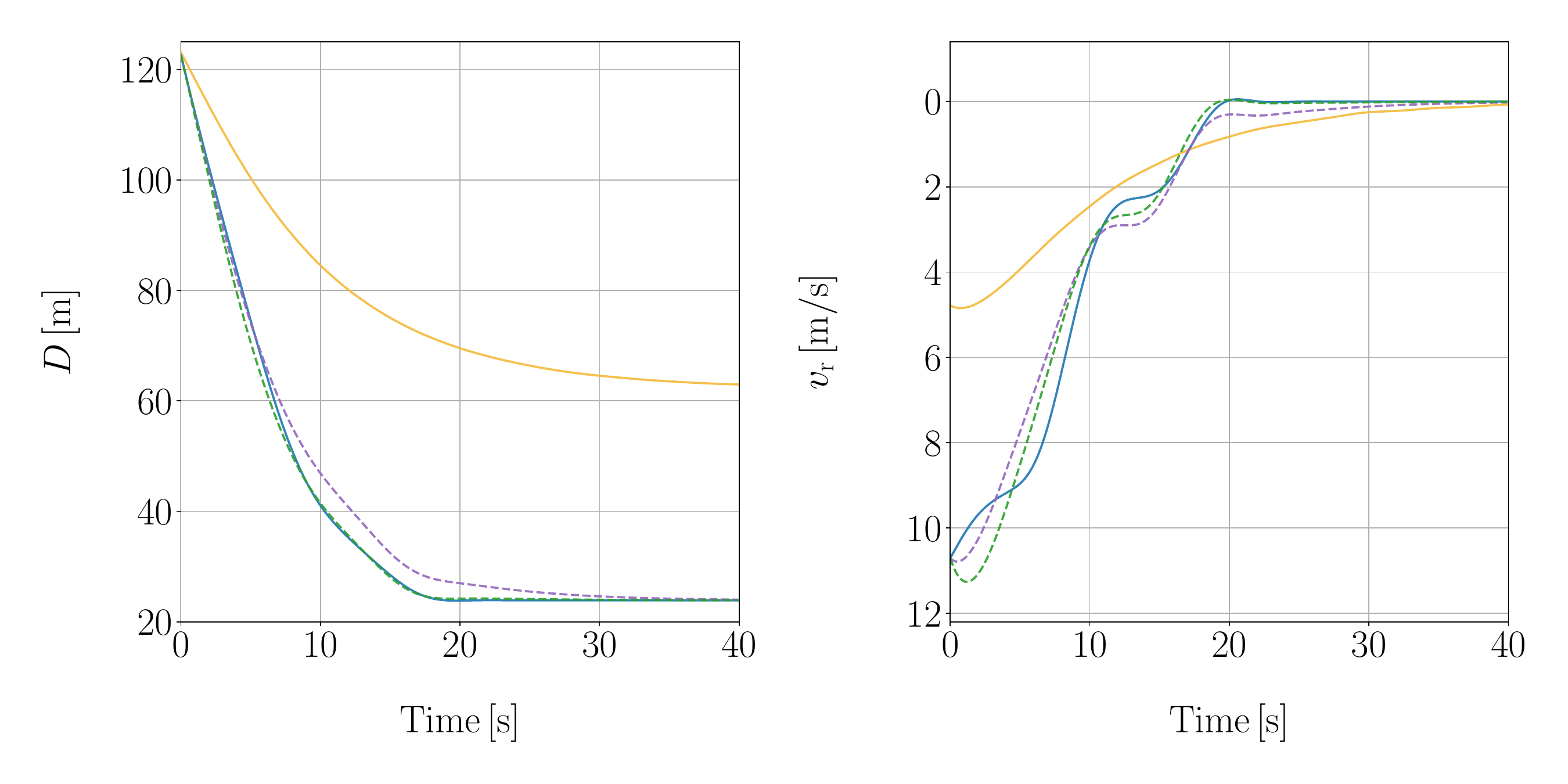}
\caption{
Transformation from 
\slowline\   $\mathcal{D}_\mathrm{slow}^\mathrm{mdl}$ to
\linearline\ $\mathcal{D}_\Delta$ and
\fmline\     $\mathcal{D}_\mathrm{FM}$
compared to a real-world fast approach scenario
\fastline\   $\mathcal{D}_\mathrm{fast}^\mathrm{meas}$.
}
\vspace{-15pt}
\label{fig:traj-drv}
\end{center}
\end{figure}

To confirm that the transported parameters are meaningful at the level of the closed-loop behavior, a generated point ($m=0$) corresponding to a trajectory for which the linear shift performs poorly was selected as an illustrative case, as shown in Fig.~\ref{fig:par-drv}, and the ACC system was simulated with the parameters obtained from each of the two transformation approaches.
Fig.~\ref{fig:traj-drv} shows the relative distance $D(t)$ and relative velocity $v_\mathrm{r}(t)$ of the corresponding approach maneuvers.

The trajectory generated using the flow matching parameter vector follows the measured fast-approach reference noticeably more closely than the trajectory obtained using the linear shift $\mathcal{D}_\Delta$.
While the linear transformation shifts every parameter vector by the same vector, flow matching learns a continuous and nonlinear transport map between the underlying parameter distributions and adapts the displacement to each individual parameter vector while preserving the coupling between the parameters.
The maximum absolute error for the relative distance $D(t)$ with respect to the recorded fast-approach maneuver amounts to $e_\mathrm{abs}^\Delta = \SI{4.43}{\meter}$ for the linear parameter shift and $e_\mathrm{abs}^\mathrm{FM} = \SI{2.03}{\meter}$ for flow matching.
Flow matching therefore reduces the maximum absolute error by approximately \SI{54}{\percent} compared with the linear transformation, consistent with the observations in Fig.~\ref{fig:traj-drv}, and shows its suitability as a behaviorally consistent scenario-to-scenario transformation.

\subsection{Systematic Evaluation of Flow Matching}
\label{subsec:quant}
In addition to the qualitative assessment of Sec.~\ref{subsec:quali}, the agreement between the generated and the measured parameter distributions is quantified by three complementary distance measures.
These were evaluated across all drivers and for a single driver.
In all evaluations and for each distance measure, the best value is highlighted in bold, while the second-best value is enclosed in a box.

\subsubsection{Distances}
\label{subsubsec:distances}
To evaluate how closely the generated parameter distribution matches the measured target distribution, three complementary measures are used.
The Wasserstein-2 distance (earth mover's distance) quantifies the full point-wise transport cost between the two empirical distributions~\cite{gabriel19:cot}.
For $N$ generated points $\{x_i\}$ and $M$ target points $\{y_j\}$ with uniform weights $a_i=1/N$ and $b_j=1/M$, it is given by the discrete optimal-transport problem
\begin{equation}
	W_2 = \left( \min_{\gamma \in \Gamma(a,b)}
	\sum_{i=1}^{N}\sum_{j=1}^{M} \gamma_{ij}\,\|x_i - y_j\|_2^2 \right)^{1/2},
\end{equation}
where $\gamma_{ij}\ge 0$ is the mass transported from $x_i$ to $y_j$ and $\Gamma(a,b)=\{\gamma\in\mathbb{R}_{\geq 0}^{N\times M}:\sum_j\gamma_{ij}=a_i,\;\sum_i\gamma_{ij}=b_j\}$ enforces that all source mass is moved and all target mass is filled.
A smaller $W_2$ indicates a closer match between the generated and the measured parameter distribution.

In addition, $\Delta\mu = \left\| \hat{\mu}_{\mathrm{gen}} - \hat{\mu}_{\mathcal{D}_\mathrm{fast}} \right\|_2$ and 
$\|\Delta\Sigma\|_F = \left\| \Sigma_{\mathrm{gen}} - \Sigma_{\mathcal{D}_\mathrm{fast}} \right\|_F$ isolate the agreement of the first two moments, i.e., the location and the shape (spread and orientation) of the distributions.

Among these measures, $W_2$ compares the complete distributional geometry rather than only its first two moments.
Thus, it is used as the primary evaluation criterion throughout the remainder of this work, while $\Delta\mu$ and $|\Delta\Sigma|_F$ serve as complementary diagnostics that attribute the residual discrepancy to location and shape, respectively.

\subsubsection{Population-averaged performance}
Table~\ref{tab:drv_all_fm} reports the three distance measures over all six drivers, approach maneuvers, and $10$ random seeds.
All measures are normalized by the corresponding values of the linear transformation, such that a value of one represents the baseline performance.
Starting from the slow-approach dataset, the transformed samples are compared with the measured fast-approach dataset.
\begin{table}[b]
\centering
\caption{Scenario-to-scenario distribution matching across drivers.}
\setlength{\tabcolsep}{3pt}
\begin{tabular}{@{}llll@{}}
\toprule
Model & $W_2^\mathrm{norm}$ & $\Delta\mu^\mathrm{norm}$ & $\|\Delta\Sigma\|_F^\mathrm{norm}$\\
\midrule
Linear & $1.00 \pm 0.00$ & $1.00 \pm 0.00$ & $1.00 \pm 0.00$ \\
FM & $0.18 \pm 0.02$ & $\tbox{0.05} \pm 0.02$ & $0.18 \pm 0.16$ \\
FM, DCC & $0.21 \pm 0.05$ & $\bf{0.05} \pm 0.03$ & $0.27 \pm 0.21$ \\
FM, DCC, DrvEmb & $0.17 \pm 0.05$ & $0.10 \pm 0.07$ & $\tbox{0.13} \pm 0.06$ \\
FM, CFG & $0.19 \pm 0.04$ & $0.07 \pm 0.04$ & $0.23 \pm 0.15$ \\
FM, CFG, DCC & $\bf{0.15} \pm 0.02$ & $0.06 \pm 0.02$ & $0.18 \pm 0.12$ \\
FM, CFG, DCC, DrvEmb & $\tbox{0.17} \pm 0.03$ & $0.09 \pm 0.06$ & $\bf{0.11} \pm 0.07$ \\
\bottomrule
\end{tabular}
\label{tab:drv_all_fm}
\end{table}

Flow matching reduces the normalized Wasserstein distance from $W^\mathrm{norm}_2=1$ (linear) to $W^\mathrm{norm}_2=0.18$, corresponding to a reduction by roughly a factor of six.
The mean deviation is likewise reduced from $\Delta\mu^\mathrm{norm}=1$ to $\Delta\mu^\mathrm{norm}=0.05$, while the covariance error decreases from $\|\Delta\Sigma\|_F^\mathrm{norm}=1$ to $\|\Delta\Sigma\|_F^\mathrm{norm}=0.18$.
These results show that flow matching more accurately reproduces both the location and the distributional shape of the fast-approach data than the linear transformation.

Conditioning on the $\beta$-VAE driving profile via CFG and combining this with DCC achieves the lowest Wasserstein distance ($W_2^\mathrm{norm}=0.15$) among all variants.
Despite its slightly higher mean deviation and covariance deviation, its lowest Wasserstein distance indicates the most accurate overall distributional match.

\subsubsection{Per-driver performance}
In addition, the performance was computed in-sample, individually per driver, with the results for Driver 5 shown exemplarily in Table \ref{tab:drv5_fm}.
\begin{table}[b]
\caption{Scenario-to-scenario distribution matching for driver 5.}
\centering
\setlength{\tabcolsep}{3pt}
\begin{tabular}{@{}llll@{}}
\toprule
Model & $W_2^\mathrm{norm}$ & $\Delta\mu^\mathrm{norm}$ & $\|\Delta\Sigma\|_F^\mathrm{norm}$\\
\midrule
Linear & $0.93 \pm 0.00$ & $0.98 \pm 0.00$ & $0.23 \pm 0.00$ \\
FM & $\bf{0.15} \pm 0.01$ & $\tbox{0.06} \pm 0.03$ & $0.24 \pm 0.05$ \\
FM, DCC & $0.16 \pm 0.01$ & $\bf{0.05} \pm 0.03$ & $0.27 \pm 0.05$ \\
FM, DCC, DrvEmb & $0.16 \pm 0.02$ & $0.07 \pm 0.03$ & $0.24 \pm 0.05$ \\
FM, CFG & $0.17 \pm 0.01$ & $0.08 \pm 0.03$ & $\tbox{0.20} \pm 0.04$ \\
FM, CFG, DCC & $0.27 \pm 0.08$ & $0.24 \pm 0.11$ & $\bf{0.18} \pm 0.09$ \\
FM, CFG, DCC, DrvEmb & $\tbox{0.16} \pm 0.02$ & $0.08 \pm 0.04$ & $0.22 \pm 0.03$ \\
\bottomrule
\end{tabular}
\label{tab:drv5_fm}
\end{table}
Across all drivers, flow matching with CFG, DCC and DrvEmb consistently ranked among the two best-performing configurations, ranging from the best result for Driver 4 ($W_2^\mathrm{norm}=0.10$) to the worst for Driver {1} ($W_2^\mathrm{norm}=0.32$).
This differs from the population results in Table \ref{tab:drv_all_fm}, where flow matching with CFG and DCC achieves the best overall performance.
Across individual drivers, however, its performance varies substantially, ranging from $W_2^\mathrm{norm}=0.17$ for Driver 4 to $W_2^\mathrm{norm}=0.55$ for Driver 3.
Note that distances over all drivers are evaluated on the pooled parameter distribution and is therefore not the mean of the driver-wise values.

\subsection{Generalization to Unseen Drivers}
\label{subsec:unseen-drivers}
To assess the ability of the learned transformation to generalize to unseen drivers, the vector field was retrained using a leave-one-driver-out cross-validation scheme.
For each driver, the vector field is retrained from scratch on all remaining drivers, with the held-out driver entirely excluded from training.
However, the $\beta$-VAE is trained once on the trajectories of all drivers, providing an optimistic performance estimate. 
The distances were then computed between the generated and measured fast-approach distributions for the held-out driver.
Variants using explicit driver embedding are not applicable in this setting, as the held-out driver has no representation in the training data.
Across all drivers, the parameters generated for Driver 5 show the closest agreement with the measured fast-approach distribution (Table~\ref{tab:drv5_fm_without_training_drv5}), whereas Driver 4 shows the poorest agreement (Table~\ref{tab:drv4_fm_without_training_drv4}).

For unseen drivers, flow matching with DCC provides more consistent performance across drivers than the other variants: for every held-out driver, it ranks among the two configurations with the lowest normalized Wasserstein distance.
This consistently competitive performance may be attributed to the driver-consistent pairing, which shapes the transport without requiring driver-specific input.
In contrast, profile-based conditioning appears to benefit more from the availability of similar training data, as the transformer-based $\beta$-VAE driving profile can then provide a more informative conditioning signal.

\begin{table}[t]
\caption{Evaluation on Driver 5 held out during training.}
\centering
\begin{tabular}{@{}llll@{}}
\toprule
Model & $W_2^\mathrm{norm}$ & $\Delta\mu^\mathrm{norm}$ & $\|\Delta\Sigma\|_F^\mathrm{norm}$\\
\midrule
Linear & $0.93 \pm 0.00$ & $0.98 \pm 0.000$ & $0.23 \pm 0.00$ \\
FM & $\tbox{0.16} \pm 0.02$ & $\tbox{0.08} \pm 0.04$ & $\tbox{0.23} \pm 0.06$ \\
FM, DCC & $\bf{0.15} \pm 0.01$ & $\bf{0.05} \pm 0.02$ & $0.25 \pm 0.05$ \\
FM, CFG & $0.23 \pm 0.05$ & $0.18 \pm 0.07$ & $\bf{0.16} \pm 0.04$ \\
FM, CFG, DCC & $0.50 \pm 0.14$ & $0.51 \pm 0.17$ & $0.27 \pm 0.13$ \\
\bottomrule
\end{tabular}
\label{tab:drv5_fm_without_training_drv5}
\end{table}

\begin{table}[t]
\caption{Evaluation on Driver 4 held out during training.}
\centering
\begin{tabular}{@{}llll@{}}
\toprule
Model & $W_2^\mathrm{norm}$ & $\Delta\mu^\mathrm{norm}$ & $\|\Delta\Sigma\|_F^\mathrm{norm}$\\
\midrule
Linear & $1.46 \pm 0.00$ & $1.56 \pm 0.00$ & $0.13 \pm 0.00$ \\
FM & $0.71 \pm 0.11$ & $0.75 \pm 0.11$ & $0.09 \pm 0.03$ \\
FM with DCC & $\tbox{0.67} \pm 0.06$ & $\tbox{0.71} \pm 0.06$ & $0.08 \pm 0.04$ \\
FM with CFG & $\bf{0.60} \pm 0.09$ & $\bf{0.64} \pm 0.09$ & $\tbox{0.06} \pm 0.04$ \\
FM with CFG, DCC & $0.76 \pm 0.07$ & $0.80 \pm 0.08$ & $\bf{0.06} \pm 0.02$ \\
\bottomrule
\end{tabular}
\label{tab:drv4_fm_without_training_drv4}
\end{table}


\section{Conclusions and Outlook}
\label{sec:conclusion}
In this work, driver-individual scenario-to-scenario parameter generation was formulated as a distribution transport problem and addressed using flow matching.
The proposed approach learns a vector field that transforms controller-parameter distributions between scenarios while capturing driver-specific characteristics.
Compared with a linear transformation, flow matching captures changes in both location and shape of the target distribution.

Evaluation on a large-scale real-world dataset demonstrated the suitability of flow matching for driver-individual parameter generation and showed improved agreement with measured target distributions.
Furthermore, a transformer-based $\beta$-VAE architecture was proposed to learn compact driving profiles from trajectory data, enabling profile-based conditioning of the transport.
While this conditioning was particularly effective for drivers represented during training, driver-consistent coupling provided more robust generalization to unseen drivers.

Future work will investigate a broad range of driving scenarios and environmental conditions, as well as transport and conditioning methods specifically tailored to driver-individual scenario transformation.
Finally, the proposed approach needs to be validated in-vehicle to evaluate its applicability to real-world driver assistance systems, including online personalization based on a limited number of observed maneuvers.

\bibliographystyle{IEEEtran}
\bibliography{paper3}

\end{document}